\documentclass[twocolumn,showpacs,preprintnumbers,amsmath,amssymb,10pt,a4paper]{revtex4}

\usepackage{geometry}
\usepackage{graphicx}
\usepackage{dcolumn}
\usepackage{bm}
\usepackage[english]{babel}

\newcommand{\be}{\begin{equation}}
\newcommand{\ee}{\end{equation}}
\newcommand{\BE}{\begin{eqnarray}}
\newcommand{\EE}{\end{eqnarray}}

\begin{document}

\preprint{}
\title{Stochastic waves in a Brusselator model with nonlocal interaction}

\author{Tommaso Biancalani}
\email{tommaso.biancalani@postgraduate.manchester.ac.uk}
\author{Tobias Galla}
\email{tobias.galla@manchester.ac.uk}
\author{Alan J. McKane}
\email{alan.mckane@manchester.ac.uk}

\affiliation{Theoretical Physics Division, School of Physics and Astronomy, 
University of Manchester, Manchester M13 9PL, United Kingdom}

\date{\today}

\begin{abstract}
We show that intrinsic noise can induce spatio-temporal phenomena such as 
Turing patterns and travelling waves in a Brusselator model with nonlocal 
interaction terms. In order to predict and to characterize these stochastic 
waves we analyze the nonlocal model using a system-size expansion. The 
resulting theory is used to calculate the power spectra of the stochastic 
waves analytically, and the outcome is tested successfully against simulations.
We discuss the possibility that nonlocal models in other areas, such as 
epidemic spread or social dynamics, may contain similar stochastically-induced 
patterns. 
\end{abstract}


\maketitle


\section{Introduction}
\label{intro}
The underlying idea of the theory of complex systems is that complex patterns
or structures can be generated from simple models; one of the key papers 
leading to this insight was the classic paper of Turing~\cite{Turing1952} on
what are now known as reaction-diffusion equations. Most of the theoretical
studies of pattern formation have followed Turing and used partial 
differential equations (pdes) to specify the model describing the 
system~\cite{Cross2009}. However there is a potential problem with this 
approach: the parameter range for which the patterns exist in the model can 
be very restricted, in contrast with what is seen in real systems. For 
instance, to observe Turing patterns in simple reaction-diffusion systems 
described by pdes requires that the diffusivities of the species be of 
different orders~\cite{Murray1993,Cross2009}. The limited range of parameters 
for which patterns are seen could be attributed to the simplicity of the 
model chosen to describe the process, however for systems with an underlying 
molecular basis another explanation has recently been put 
forward~\cite{Butler2009,Biancalani2010,Scott2011}. These authors have
observed that Turing-like patterns exist for a much greater range of parameter 
values if the discrete nature of the molecules comprising the system is taken
into account. The resulting patterns are sometimes referred to as stochastic 
Turing patterns~\cite{Biancalani2010} or quasi-Turing 
patterns~\cite{Butler2011}, and they may be analyzed using the theory of 
stochastic processes, appropriate given the noise is created as a consequence 
of the discreteness of the system.

In this paper we investigate a model which not only shows Turing and 
stochastic Turing patterns, but also travelling waves and stochastic travelling
 waves (referred to as stochastic waves in the following), the latter having 
the same relation to travelling waves as Turing patterns have to stochastic 
Turing patterns. One interesting aspect of stochastic waves is that they can 
clearly be seen in computer simulations of the model. In contrast, while there 
is unambiguous evidence for the existence of stochastic Turing patterns from, 
for instance, the form of the power-spectrum of the fluctuations, direct 
visual evidence is less clear due to the noisy nature of the patterns. The 
model we study is a Brusselator with a non-local interaction term; this choice 
is largely made on the grounds that the model is simple, and so allows the 
effect to be clearly demonstrated. The non-locality seems to be an important 
ingredient in finding stochastic waves, similar to the observation that 
travelling waves disappear in deterministic models when interactions are made 
more local~\cite{Nicola2006}. Travelling waves have been observed in chemical 
reaction systems in~\cite{zaikin1970,Field1985,swinney1987}, and also in 
other types of population-based 
systems~\cite{mayt1991,ranta1997,grenfell2001,cummings2004}.

The description of patterns through an analysis of pdes of the 
reaction-diffusion type is well-developed~\cite{Cross2009}, but the analogous 
stochastic systems have received very little attention. The common starting
point of the work that has been carried 
out~\cite{Butler2009,Biancalani2010,Scott2011} has been the master equation
(continuous-time Markov chain), although the details of the techniques used
to analyze this equation have differed. However the whole idea of 
stochastic patterns, and the methods which may be used to analyze them, follows 
on from the idea of stochastic cycles, or quasi-cycles, and their detailed 
study over the last few years. Since the essential idea behind stochastic 
patterns can be clarified with reference to the well-studied mechanism behind 
stochastic cycles, we turn to their description.

The context in which stochastic cycles appear is in the relation between 
stochastic individual-based models and the corresponding deterministic 
descriptions of their dynamics. Interacting many-particle systems are typically 
defined by a set of stochastic rules at the microscopic level. Such systems 
are common in chemistry and in biology~\cite{Kampen1997}, but they are also 
used to model stochastic dynamics in 
epidemiology~\cite{Alonso2006, Rozhnova2009}, in population 
dynamics~\cite{McKane2005, Traulsen2005} and in the social 
sciences~\cite{Castellano2009}. Frequently in formulating descriptions of 
these systems intrinsic noise, due to the discrete nature of the constituents, 
is treated as a negligible perturbation to the dominant deterministic dynamics.
It has been known for many years that neglecting such fluctuations is not 
always justified, on the contrary, intrinsic noise can fundamentally change 
the temporal evolution of these systems. The activity in this area over the 
last few years is due to the realization that a method imported from 
statistical physics --- the van Kampen system-size 
expansion~\cite{Kampen1997} --- can be used to \textit{quantitatively} 
understand the deviations to the deterministic dynamics caused by these 
stochastic effects. Effectively one uses some measure of the inverse system 
size, for instance its inverse volume $V^{-1}$, as a perturbation parameter 
to investigate the stochastic dynamics. 

If truncated after the lowest order, the system-size expansion provides a 
systematic path from stochastic microscopic interacting-particle models to 
a deterministic description in terms of differential equations. This 
truncation effectively corresponds to taking the thermodynamic limit, 
i.e.~to considering the limiting case of infinite systems. If carried out 
to next-to-leading order the system-size expansion can successfully 
characterize stochastic effects in the limit of large, but finite systems.
Of particular interest is the case where the deterministic model has a 
stable fixed point, which is a spiral, so that perturbations away from the 
fixed point decay in an oscillatory manner. The effect of the stochasticity 
is to excite this decaying mode and create sustained oscillations, which have 
an amplitude larger than might naively be expected --- stochastic 
amplification~\cite{McKane2005}. These are the stochastic cycles. 

In this description of stochastic cycles no mention has been made of the spatial
degrees of freedom, and indeed most of the studies of stochastic cycles have 
been confined to well-mixed systems where spatial effects are ignored. However,
in a similar way, intrinsic fluctuations can excite more complicated 
spatio-temporal modes in spatially extended systems, as seen for example in 
a predator-prey model of the Volterra type~\cite{lugo2008}. In this 
spatially-extended model the intrinsic noise leads to spatially uniform 
temporal oscillations, i.e.~at zero wavenumber. Stochastic Turing patterns, 
just like stochastic cycles, are triggered by internal fluctuations, but they 
occur at non-zero wavenumber. To date they have been found in a spatial version 
of the Levin-Segel predator-prey model~\cite{Butler2009}, in the spatial 
Brusselator model~\cite{Biancalani2010} and in a model of embryonic 
patterning~\cite{Scott2011}. What does not change is the ability of the 
system-size expansion to accurately predict the features of the excitations.

The main purpose of the present paper is to show that in addition to uniform 
spatial oscillations and time-independent Turing patterns, intrinsic noise 
can also trigger travelling waves, i.e.~phenomena which are both spatial and 
temporal simultaneously. To show that such effects may occur we study a 
variation of the Brusselator model with a nonlocal interaction term, mediated 
by an exponential kernel in space. We believe that the existence of stochastic 
travelling waves is widespread, and we choose the Brusselator model because it
is probably the most widely used model to illustrate such phenomena, and so
is familiar to a wide number of researchers in the field. A disadvantage
is that nonlocal interactions which seem to be required to see the effect are
less easy to motivate in a chemical model such as the Brusselator, but we 
believe that this is outweighed by the advantage of using a simple model to
illustrate the idea. So while several physical mechanisms for nonlocal 
interactions are given in~\cite{Nicola2006}, and in biological and 
social systems such effects are far easier to describe and motivate, our 
desire in this paper to avoid specific mechanisms and extraneous details,
leads us to formulate the nonlocal interaction in a simple and generic way.

We show that the presence of such a nonlocal term promotes travelling waves, 
and that stochasticity can bring about stochastic travelling waves in 
situations where the deterministic system has a stable uniform fixed point. 
All this is carried out analytically through use of the system-size 
expansion --- extended to deal with the non-local interaction --- and checked 
numerically using the Gillespie algorithm~\cite{Gillespie1976,Gillespie1977}. 

The remainder of the paper is structured as follows. In Section~\ref{sec:model} 
we introduce the model and discuss its behavior on the deterministic 
level. In Section~\ref{sec:stoch_anal} we carry out a detailed analysis of the 
stochastic dynamics by means of an expansion in the inverse system size. 
Section~\ref{sec:results} then contains our main results, including an 
analytical characterization of stochastic waves and confirmation through 
numerical simulations. Our findings are summarized in Section~\ref{sec:concl} 
where we also discuss possible future lines of research. Two appendices contain 
mathematical details: the first a summary of the conditions under which
Turing and wave instabilities occur and the second aspects of the calculations 
and results from the system-size expansion.

\section{Model definition}
\label{sec:model}
The stochastic model is defined as a collection of uniform cells, indexed 
by $i$, each of which has volume $V$. For convenience these can be thought of
as cubes in three dimensions, but other regular shapes in other dimensions can 
also be considered. In fact, since our main aim here is to illustrate the 
idea of stochastic waves and their characterization, our results will largely 
apply to a one-dimensional system, i.e. a chain of cells. Simulations are then 
less time-consuming, but the model still exhibits all features we are 
interested in studying. Generalization of the analytic results to 
higher-dimensional models is straightforward. Unlike previous 
work~\cite{lugo2008}, we will assume that the number of cells is infinite, 
that is, the underlying space is of infinite extent. This avoids technical 
complications when nonlocal interactions are present.

In every cell molecules of two species, $X$ and $Y$, interact through the 
reactions of the Brusselator model~\cite{Prigogine1971}:
\begin{equation} 
\label{spbrus}
\begin{split}
\quad \emptyset _i &\overset{a}{\rightarrow} X_i, \\
\quad X_i &\overset{b}{\rightarrow} Y_i, \\
\quad 2X_i + Y_i &\overset{c}{\rightarrow} 3X_i, \\
\quad X_i &\overset{d}{\rightarrow} \emptyset _i.
\end{split}
\end{equation}
The rates at which the reactions occur are denoted by $a, b, c$ and $d$, and
$X_i$ and $Y_i$ are molecules which are in cell $i$ at the time the reaction 
occurs. The third reaction may occur between an $X$-molecule in cell $i$ and a 
$Y$-molecule in any other cell (with rates depending on the distance between 
the two cells), but the effect is to reduce the number of $Y$ molecules in cell
$i$ by $1$. This constitutes the nonlocal interaction; other choices are 
possible, but this was made on grounds of simplicity. The notation used to 
describe the chemical types and the rates in the Brusselator model in the 
literature is not standard, but we follow most closely~\cite{Boland2009}. The 
number of $X$ and $Y$ molecules in cell $i$ will be denoted by $n_i$ and $m_i$ 
respectively. We will also use $\mathbf n$ and $\mathbf m$ to represent the 
spatial vectors with components $n_i$ and $m_i$ respectively.

The transition rate from the state $(\mathbf n, \mathbf m)$ to the state
$(\mathbf{n}', \mathbf{m}')$ will be denoted by
$T(\mathbf{n}', \mathbf{m}' | \mathbf{n}, \mathbf{m})$, but to lighten the
notation we will only list the variables which have changed in a given 
reaction. These functions are found by invoking mass action:
\begin{equation} 
\label{spacetransprob}
\begin{split}
T_{1}(n_i+1,m_i|n_i,m_i) &= \; a, \\
T_{2}(n_i-1,m_i+1|n_i,m_i) &= b \; \frac{n_i}{V}, \\
T_{3}(n_i+1, m_i-1|n_i,m_i) &= c \; \frac{{n_i}^2}{V^2} \; 
\Lambda \sum_{j=-\infty}^{\infty} e^{-\sigma |i-j|} \; \frac{m_{j}}{V}, \\
T_{4}(n_i-1, m_i|n_i,m_i) &= d \; \frac{n_i}{V},
\end{split}
\end{equation}
where the subscripts on the rates refer to the four reactions in 
Eq.~(\ref{spbrus}). These transition rates are as in the usual Brusselator
model~\cite{Boland2009}, except for the third which has the nonlocal 
character mentioned earlier. We stress again that it is the influence of the
$Y$ molecule in cell $j$ that causes the reaction, but that the effect is in
cell $i$ which is at a distance $|i-j|$ away. The effect is to increase the 
number of $X$ molecules in cell $i$ by one and to decrease the number of 
$Y$ molecules in cell $i$ by one. The form of interaction was taken to be
exponential because this is a frequent choice, but once again on grounds of
simplicity. The constant $\sigma$ expresses the range of the interaction and 
$\Lambda$ is a normalization constant whose choice will be discussed later. 
As a technical aside, the third transition rate should also have a factor of 
$\theta(m_i)$ --- the Heaviside step function --- present to prevent the 
number of $Y$ molecules in cell $i$ becoming negative, but given we will be 
looking at a regime far from $m_i = 0$, this condition is irrelevant.

In addition to the reactions given in Eq.~(\ref{spbrus}), ``migration'' 
reactions which describe molecular diffusion from one cell to another, have 
to be specified. For a given cell $i$, molecules of the two species $X$ and 
$Y$ may diffuse into or diffuse out of a neighboring cell $j$:
\begin{equation} 
\label{migrreaz}
\begin{split}
X_i &\overset{\alpha}{\rightarrow} X_j, \quad X_j 
\overset{\alpha}{\rightarrow} X_i, \\
Y_i &\overset{\beta}{\rightarrow} Y_j, \quad Y_j 
\overset{\beta}{\rightarrow} Y_i. \\
\end{split}
\end{equation}
These reactions have transition rates given by
\begin{equation} 
\label{migrprob}
\begin{split}
T_{5}(n_i-1, n_j+1|n_i,n_j) &= \alpha \; \frac{n_i}{V z}, \\
T_{6}(n_i+1, n_j-1|n_i,n_j) &= \alpha \; \frac{n_j}{V z}, \\
T_{7}(m_i-1,m_j+1|m_i,m_j) &= \beta \; \frac{m_i}{V z}, \\
T_{8}(m_i+1, m_j-1|m_i,m_j) &= \beta \; \frac{m_j}{V z}.
\end{split}
\end{equation}
Here $z$ is the number of nearest neighbors of a cell and the index $j$ denotes
a nearest neighbor of cell $i$. For the one-dimensional model $z=2$ and 
$j\in \{ i-1, i+1\}$. It is also worth remarking that we have not imposed a
fixed limit on the number of molecules permitted in a cell (in contrast with
some previous work~\cite{lugo2008,Biancalani2010}; this is reflected in the
fact we use the inverse volume, $V^{-1}$, rather than the inverse total number
of molecules, as the expansion parameter).

Since we have assumed that the transition rates only depend on the current 
state of the system, the stochastic process is Markov, and so the probability 
distribution for the system being in state $(\textbf{n},\textbf{m})$ at time
$t$, $P_{\textbf{n},\textbf{m}}(t)$, satisfies the master 
equation~\cite{Kampen1997}:
\BE 
\label{spme}
{d \over d t}P_{\textbf{n},\textbf{m}}(t) &=& 
\sum_{{(\mathbf n', \mathbf m') \ne \choose (\mathbf n, \mathbf m)}}  
\Biggl [ T(\mathbf n,\mathbf m|\mathbf n',\mathbf m')
P_{\textbf{n}',\textbf{m}'}(t) \nonumber\\
&-& T(\mathbf n',\mathbf m'|\mathbf n,\mathbf m)
P_{\textbf{n},\textbf{m}}(t)\Biggr].
\EE
In what follows we will write discrete variables as subscripts and continuous 
variables as arguments of functions (such as in $P_{\textbf{n},\textbf{m}}(t)$).
The sole exception is for the transition rates 
$T(\mathbf{n}', \mathbf{m}' | \mathbf{n}, \mathbf{m})$, where for the sake of
clarity we have written the discrete variables as arguments of the function $T$.

We will discuss the analysis of the master equation (\ref{spme}) using the 
system-size expansion in the next section. For the rest of this section we will
obtain the deterministic equation which holds in the limit $V \to \infty$
directly, without using the system-size expansion, and then study the stability
of the homogeneous state.

We begin by defining the concentrations of the $X$ and $Y$ molecules in 
cell $i$ in the infinite volume limit by
\be 
\phi_i(t) = \lim_{V \to \infty} \frac{\langle n_i \rangle}{V} \ \ , \ \ 
\psi_i(t) = \lim_{V \to \infty} \frac{\langle m_i \rangle}{V},
\label{phi_psi}
\ee
where $\langle \ldots \rangle$ is the mean with respect to the probability
distribution $P_{\textbf{n},\textbf{m}}(t)$. Multiplying Eq.~(\ref{spme}) by 
$n_i$ and $m_i$ respectively, and using the fact that in the deterministic 
limit the probability distribution is a delta-function so that 
$\langle n^{\ell}_{i} \rangle = \langle n_i \rangle^{\ell}$ etc, one finds 
the following ordinary differential equations:
\BE
\label{deter}
 {d \over d\tau} \phi_i &=& a -(b+d)\phi_i \nonumber \\ &&+ 
c \:\phi_i^2 \:\Lambda \sum_{j=-\infty}^{\infty}  e^{-\sigma |j|} \:\psi_{i-j} + 
\alpha \:\Delta \phi_i, \nonumber \\
{d \over d\tau} \psi_i &=& b \: \phi_i - 
c \: \phi_i^2 \:\Lambda  \sum_{j=-\infty}^{\infty}  e^{-\sigma |j|} 
\:\psi_{i-j} \nonumber \\ 
&&+ \beta \:\Delta \psi_i,
\EE
where $\tau = t/V$ is a rescaled time and where 
$\Delta f_i = f_{i+1} - 2f_{i} + f_{i-1}$ is the discrete one-dimensional 
Laplacian.

We choose the normalization constant $\Lambda$ so as to satisfy
\begin{equation} 
\label{norm}
\Lambda \sum_{j=-\infty}^{\infty}  e^{-\sigma |j|} = 1.
\end{equation}
By doing so, the deterministic equations \eqref{deter} have the homogeneous 
solution: 
\begin{equation} 
\label{fps}
\phi_i=\phi^* = \dfrac{a}{d}, \qquad \psi_i=\psi^* = \dfrac{b d}{a c},
\end{equation}
which are the same as those of the conventional Brusselator model (obtained 
from our model by replacing the non-local interaction by a local term). The 
expression for $\Lambda$ can be summed to yield
\begin{equation} 
\label{lambda}
\Lambda = \frac{e^{\sigma}-1}{e^{\sigma}+1}.
\end{equation}

Eqs. (\ref{deter}) are a set of reaction-diffusion equations of the type 
usually defining the starting point for finding Turing patterns. The analysis 
starts by examining if the homogeneous solution (\ref{fps}) is unstable to 
spatially inhomogeneous small perturbations. To do this we introduce small 
perturbations 
\be
\delta\phi_i(t) = \phi_i(t) - \phi^{*} \ \ , \ \ 
\delta\psi_i(t) = \psi_i(t) - \psi^{*},
\label{small}
\ee 
into the deterministic equations (\ref{deter}) and keep only linear terms in 
$\delta\phi_i(t)$ and $\delta\psi_i(t)$. This gives
\BE
\label{linearize}
{d \over d\tau} \delta \phi_i &=& -(b+d) \delta \phi_i + 
c \:\phi^{*2} \:\Lambda \sum_{j=-\infty}^{\infty}  e^{-\sigma |j|} \: 
\delta \psi_{i-j} \nonumber \\
&&+ 2\: c\: \phi^* \psi^* \delta \phi_i + \alpha \:\Delta \:  \delta \phi_i, \\
{d \over d\tau} \delta \psi_i &=& b \: \delta \phi_i - c \: 
\phi^{*2} \:\Lambda  \sum_{j=-\infty}^{\infty}  e^{-\sigma |j|} \: 
\delta \psi_{i-j} \nonumber \\
&&- 2\: c\: \phi^* \psi^*  \delta \phi_i + \beta \:\Delta \: \delta \psi_i.
\EE

The structure of these equations makes it clear that they will simplify
considerably if we go over to a Fourier representation. We therefore 
introduce the spatial Fourier transform for the infinite discrete system of 
cells:
\begin{equation} 
\label{spfourier}
\tilde f(k) = \sum_{j=-\infty}^{\infty} e^{-\mathrm i j k} \: f_j, 
\quad  f_j = \frac{1}{2\pi}\int^{2\pi}_{0}dk\, e^{\mathrm i k j} \: \tilde f(k).
\end{equation}
Note that $k$ is a continuous variable which takes values in the first 
Brillouin zone, $[0,2\pi]$. Fourier transforming the equations 
\eqref{linearize} gives:
\BE 
\label{fmf}
{\partial \over \partial\tau}  \delta \tilde{\phi} &=& 
-(b+d) \delta \tilde\phi + c \:\phi^{*2} \:\Lambda \: 
\tilde{\mathrm e}(k) \: \delta \tilde\psi \nonumber \\
&&+ 2\: c\: \phi^* \psi^* \delta \tilde\phi + \alpha \:\tilde\Delta \:  
\delta \tilde\phi, \\
{\partial \over \partial\tau} \delta\tilde\psi &=& 
b \: \delta \tilde\phi - c \: \phi^{*2} \:\Lambda \: \tilde{\mathrm e}(k) \: 
\delta \tilde\psi \nonumber \\
&&- 2\: c\: \phi^* \psi^*  \delta \tilde\phi + \beta \:\tilde\Delta  \: 
\delta \tilde\psi,
\EE
where $\delta\tilde\phi = \delta \tilde{\phi}(k,\tau)$ and similarly for
$\delta\tilde\psi$. The two functions $\tilde{\mathrm e}(k)$ and 
$\tilde\Delta\equiv\tilde\Delta(k)$ are respectively the Fourier transform 
of the exponential function and of the Laplacian, i.e.
\begin{equation}
\begin{split}
\widetilde{\Delta f}(k) &= \sum_{j=-\infty}^{\infty} e^{- \mathrm i k j} 
\left( f_{i+1} -2 f_i + f_{i-1} \right) = \\
& = 2 \left[ \cos (k) - 1\right] \tilde f(k) 
\Rightarrow \tilde\Delta \equiv  2 \left[ \cos (k) - 1\right], \\
\tilde{\mathrm e}(k) &= \sum_{j=-\infty}^{\infty} e^{-\sigma |j|} \: 
e^{- \mathrm i k j} = 
\frac{ \mbox{sinh}(\sigma) } {\mbox{cosh}(\sigma) - \cos (k)}.
\end{split}
\label{e_and_Delta}
\end{equation}

The system \eqref{fmf} may be written in a more compact form:
\begin{equation} 
\label{Jacobiandef}
{\partial \over \partial\tau} \binom{\delta \tilde\phi}{\delta \tilde\psi} = 
\mathcal J^*(k) \cdot \binom{\delta \tilde\phi}{\delta \tilde\psi}
\end{equation}
where
\be  
\label{Jacobian}
\begin{split}
&{\mathcal J}^* (k) = \\
&\left( 
\begin{array}{cc}
 -(b+d)+2\:c\:\phi^*\psi^* + \alpha \: \tilde\Delta  & c \: \Lambda \: 
\phi^{*2} \: \tilde {\mathrm e}(k) \\
 b-2\:c\:\phi^*\psi^* & -c \:\Lambda\: \phi^{*2} \: 
\tilde{\mathrm e}(k)  +\beta \: \tilde\Delta
\end{array} 
\right).\\
\end{split}
\ee

The eigenvalues of the Jacobian matrix (\ref{Jacobian}), $\lambda_1(k)$ and 
$\lambda_2(k)$, yield information about whether perturbing the homogeneous 
solution leads to pattern formation. If both $\lambda_1$ and $\lambda_2$ have 
negative real part 
(i.e. $\mbox{Re} [\lambda_i(k)] < 0, \; \forall k, \; i=1,2$), the homogeneous 
state is stable: every perturbation will eventually die out and no pattern will 
develop. If, on the other hand, there is an eigenvalue at a non-zero $k$ with
a positive real part, then a spatially modulated instability occurs: a 
perturbation will grow in magnitude taking the system from the homogeneous 
state to one with wavenumber defined by $k$. This growth will eventually be 
saturated by the non-linear terms leading to a pattern of characteristic wave 
number $k$.

This linear analysis of the homogeneous state is also able to determine whether 
the resulting pattern is steady or oscillatory, by looking at the imaginary 
part of the eigenvalues $\omega_i \equiv \mbox{Im}[\lambda_i]$. Steady patterns 
correspond to  $\mbox{Im}[\lambda_i(k)] = 0$, for all unstable modes $k$, the 
case in which the instability is called a ``Turing instability''. When 
$\mbox{Im}[\lambda_i(k)] \ne 0$, for an unstable mode at a non-zero $k$, the 
system is said to undergo a ``wave instability'' as the resulting pattern will 
consist of travelling waves~\cite{Cross2009}. 

When multiple instabilities occur simultaneously, say of the Turing type for 
some $k_1$ and the wave type for some $k_2$, it becomes harder to predict if 
the final pattern will be steady or oscillatory. However, for model parameters
sufficiently close to the region in which the homogeneous state is stable,
only a single instability occurs. We will therefore study the instabilities 
at the border of the stable region, saying that there is a Turing instability
(resp.~wave instability) when the instability near the border is of the Turing
type (resp.~wave type). The criteria we use to determine the stability 
boundaries are discussed in Appendix A.

The above analysis when applied to Eq.~(\ref{Jacobian}) is discussed in
Section \ref{sec:results}. However, as explained in the Introduction we 
wish to go beyond this deterministic approximation, and examine patterns 
which emerge from stochastic effects which are a consequence of the 
underlying discreteness of the system. We therefore now go on to discuss
the stochastic analysis of the model, before collecting results on the 
determinstic and stochastic regimes in Sec.~\ref{sec:results}.
 
\section{Stochastic Analysis}
\label{sec:stoch_anal}
The analysis of the model, without making the deterministic approximation,
begins by writing down the master equation~(\ref{spme}) in a form which is 
more amenable to application of the system-size expansion. This is done by
introducing step-operators~\cite{Kampen1997}:
\begin{equation} 
\label{sstepop}
\begin{split} 
\epsilon_{X,i}^{\pm} f(\mathbf n, \mathbf m) 
&= f(\ldots, n_i\pm 1, \ldots,\mathbf m),\\ 
\epsilon_{Y,i}^{\pm} f(\mathbf n, \mathbf m) 
&= f(\mathbf n, \ldots, m_i\pm 1, \ldots).
\end{split}
\end{equation}
For example, the term in the master equation which involves the first 
reaction in Eq.~(\ref{spbrus}), and so corresponds to the function $T_1$ in 
Eq.~(\ref{spacetransprob}), would be given by
\BE
& & \sum_{i} (\epsilon_{X,i}^- -1)\;T_1(n_i+1, m_i |n_i, m_i) 
P_{\textbf{n},\textbf{m}}(t) \nonumber \\
& & = \sum_{i} (\epsilon_{X,i}^- -1)\;a\,P_{\textbf{n},\textbf{m}}(t).
\label{firstterm}
\EE
The master equation rewritten in this way, with all eight terms present, is
given by Eq.~(\ref{spme2}) in Appendix B.

We will now carry out the system-size expansion. It is important to 
realize that, for our model, this is an expansion in powers of $V^{-1/2}$, 
where $V$ is the volume per cell. The expansion therefore captures stochastic 
effects at large, but finite cell volumes. This limit should not be confused 
with the limit of a infinite number of cells. When we use the term 
`system-size expansion', we always refer to an expansion in the size (volume) 
of a single cell. Even if the volume per cell is finite, the total number of 
molecules in the system (summed over all cells) may well be infinite, if the 
system is composed of an infinite number of cells.

The system-size expansion itself involves making the time-dependent change 
of variables 
$(\mathbf n, \mathbf m) \mapsto (\boldsymbol \xi, \boldsymbol \eta)$:
\begin{equation} 
\label{spansatz}
\mathbf n \mapsto V \boldsymbol \phi (t) + \sqrt V \boldsymbol \xi,  \;
\mathbf m \mapsto V \boldsymbol \psi (t) + \sqrt V \boldsymbol \eta,
\end{equation}
where $\boldsymbol \phi (t)$ and $\boldsymbol \psi (t)$ are two time-dependent 
vectors defined by Eq.~(\ref{phi_psi}), each of these vectors having as many 
components as there are different cells in the system. From the leading-order 
term in the expansion one finds that these quantities satisfy the  
deterministic equation (\ref{deter}). 

We now change the degrees of freedom of the stochastic system to 
$\boldsymbol{\xi}$ and $\boldsymbol{\eta}$ and consider the probability 
distribution in terms of these variables as
\be
\Pi (\boldsymbol \xi, \boldsymbol \eta, t ) = P_{\mathbf n, \mathbf m}(t).
\ee
The left-hand side of the master equation (\ref{spme}) has the 
form~\cite{Kampen1997}
\begin{equation}
\label{vkfs}
\frac{dP}{dt} = \partial_t \Pi 
-\sqrt V \:\nabla_{\boldsymbol \xi}\Pi \cdot \partial_t \boldsymbol \phi 
-\sqrt V\:\nabla_{\boldsymbol \eta}\Pi \cdot \partial_t \boldsymbol\psi,
\end{equation}
where $\partial_t = \partial /\partial t$. 

To determine the nature of the expansion of the right-hand side of the master 
equation we begin with the form \eqref{spme2} given in terms of the step 
operators. This is because the step operators have a natural expansion in 
$V^{-\frac{1}{2}}$ given in Eq.~(\ref{stepdevelop}) of Appendix B. The 
introduction of a rescaled time, $\tau = t/V$ brings the master equation into 
the general form (see Appendix B)
\begin{equation}
\begin{split}
\frac{1}{V} &\partial_\tau \Pi 
-\frac{1}{\sqrt V} \left(\nabla_{\boldsymbol \xi}\Pi \cdot 
\partial_\tau \boldsymbol \phi -\:\nabla_{\boldsymbol \eta}\Pi \cdot 
\partial_\tau \boldsymbol\psi \right) \\
&= \left\{ -\frac{1}{\sqrt V} 
\left[ \boldsymbol f(\boldsymbol \phi,\boldsymbol \psi) \cdot 
\nabla_{\boldsymbol \xi} +  \boldsymbol  g(\boldsymbol \phi,\boldsymbol \psi) 
\cdot \nabla_{\boldsymbol \eta} \right]  + \frac{\mathbf L}{V} \right\} \Pi,
\end{split}
\label{gen_form}
\end{equation}
where $\mathbf L$ is a linear operator containing various derivatives in 
$\eta$ and $\xi$ and $\boldsymbol f$ and $\boldsymbol g$ are functions of
$\boldsymbol \phi$ and $\boldsymbol \psi$. 

It is now possible to match terms on both sides of the transformed master 
equation (\ref{gen_form}). The order $1/\sqrt{V}$ contributions lead to
\BE
{d \over d\tau} \phi_i &=& f_i(\boldsymbol \phi,\boldsymbol \psi), 
\nonumber \\
{d \over d\tau} \psi_i &=& g_i(\boldsymbol \phi,\boldsymbol \psi),
\label{gen_deter}
\EE
which are just the deterministic equations for the concentrations in cell 
$i$ derived in a more direct fashion earlier, and given explicitly by 
Eq.~(\ref{deter}). Matching the order $1/V$ contributions leads to an 
equation for the probability distribution $\Pi$ which describes the 
fluctuations:
\begin{equation}
\label{fp}
\begin{split}
\partial_\tau \Pi(\boldsymbol \xi, \boldsymbol \eta, t) &= 
\mathbf L \: \Pi(\boldsymbol \xi, \boldsymbol \eta, t). \\
\end{split}
\end{equation}

We have analyzed the leading order result (\ref{gen_deter}) in 
Section~\ref{sec:model}; our aim in this Section is to study the stochastic
corrections given by Eq.~(\ref{fp}). The explicit form of the 
operator $\mathbf L$ given in Appendix B shows Eq.~(\ref{fp}) to be a 
Fokker-Planck equation describing a linear stochastic process. It is more 
convenient for our purposes to use the equivalent description of the process 
in terms of a linear stochastic differential equation, or Langevin equation, 
since this allows us to take the spatial and temporal Fourier transforms. 
If we carry out a Fourier transform only with respect to the spatial variable, 
one has the form~\cite{Gardiner1985, Risken1989}
\be
\frac{\partial \tilde{\boldsymbol{\zeta}}}{\partial\tau} = 
\tilde{\mathcal{A}}(\tilde{\boldsymbol{\zeta}})
+ \tilde{\boldsymbol{\mu}}(k,\tau),
\label{FT_Langevin}
\ee
where $\tilde{\boldsymbol{\mu}}(k,\tau)$ is a Gaussian noise with zero mean and 
correlator
\be
\langle \tilde{\boldsymbol{\mu}}(k,\tau) 
\tilde{\boldsymbol{\mu}}^{\rm T}(k',\tau') \rangle =
\tilde{\mathcal{B}}(k)\,2\pi\,\delta(k-k')\delta(\tau-\tau'),
\label{FT_noise}
\ee
and where we have introduced the convenient notation
$\boldsymbol{\zeta} = (\boldsymbol{\xi}, \boldsymbol{\eta})$.

The explicit forms for $\tilde{\mathcal{A}}$ and $\tilde{\mathcal{B}}$ are 
given in Appendix B. Since the stochastic process is linear, 
$\tilde{\mathcal{A}}$ is a linear function of $\tilde{\boldsymbol{\zeta}}$ 
and $\tilde{\mathcal{B}}$ is independent of it. Both $\tilde{\mathcal{A}}$ and 
$\tilde{\mathcal{B}}$ are explicit functions of $\tau$ through their 
dependence on the solutions of the deterministic equations 
$\boldsymbol{\phi}(\tau)$ and $\boldsymbol{\psi}(\tau)$. However we are 
interested in fluctuations about the homogeneous state, and so we may take its 
values for these solutions. In this case both $\tilde{\mathcal{A}^{*}}$ and 
$\tilde{\mathcal{B}^{*}}$ lose their explicit time dependence. If the noise 
term $\tilde{\boldsymbol{\mu}}$ is omitted from Eq.~(\ref{FT_Langevin}) 
(effectively by taking $\mathcal B = 0$; see Eq.~(\ref{FT_noise})), then 
$\tilde{\boldsymbol{\zeta}}$ is nothing else than a deterministic 
perturbation of the deterministic dynamics. So it is not surprising that 
$\tilde{\mathcal{A}^{*}}$ is related to the elements of the Jacobian in the 
homogeneous state given by Eq.~(\ref{Jacobian}):
\be
\tilde{\mathcal{A}}^{*}_{r}(k,\tau) = \sum^{2}_{s=1} {\mathcal J}_{rs}^{*}(k) 
\tilde{\zeta}_{s}(k,\tau).
\label{cal_A}
\ee
Of course $\tilde{\mathcal{A}}$ retains an implicit dependence on time through 
$\tilde{\boldsymbol{\zeta}}(k,\tau)$. The form for $\tilde{\mathcal{B}}(k)$ is 
given by Eq.~(\ref{FT_cal_B}). The Langevin equation (\ref{FT_Langevin}) can 
now be solved by taking the temporal Fourier transform and determining the 
Fourier-transformed fluctuations $\tilde{\xi}(k,\omega)$ and 
$\tilde{\eta}(k,\omega)$~\cite{lugo2008}. 

For parameter values for which the homogeneous state is stable no pattern 
arises in the deterministic description. However, as discussed in the 
Introduction, we can ask if in this region of parameters the spatial system 
exhibits ordered structures once fluctuations are taken into account. To 
probe this possible fluctuationally-induced order, a useful tool is the power 
spectrum of the fluctuations about the homogeneous state, defined by
\be
P_{X}(k,\omega) = 
\left\langle \left|\tilde{\xi}(k, \omega) \right|^{2} \right\rangle \ , \ 
P_{Y}(k,\omega) = 
\left\langle \left| \tilde{\eta}(k, \omega) \right|^{2} \right\rangle .
\label{PS_defn}
\ee

In absence of order the spectra will show an almost flat profile. If instead 
some type of order is present, the power spectra will typically have a 
characteristic peak. The position of the peak in combined $(k,\omega)$-space 
determines the type of structure that is present. For example, a global 
oscillation in time will correspond to a peak of $P(k,\omega)$ at a non-zero 
value of $\omega$ and at $k=0$, whereas the power spectrum peaks at $\omega=0$ 
and at a non-zero value of $k$ for stochastic Turing patterns. As we are 
looking for stochastic waves we shall seek parameter values for which the 
power spectra display a peak at values of $(k,\omega)$ where both $k\neq 0$ and 
$\omega\neq 0$. The spectra are found both through simulations and 
analytically from the form of $\tilde{\xi}(k,\omega)$ and 
$\tilde{\eta}(k,\omega)$, derived from Eq.~(\ref{FT_Langevin}). The analytical 
calculations required to determine the power spectra are similar to those 
discussed in~\cite{lugo2008}, but with different forms for the matrices 
${\cal J}^*(k)$ and $\tilde{\cal B}^*(k)$. One finds 
\BE
\label{eq:pwspec}
P_X(\omega, k) &=& \frac{{\cal C}_{X} + 
\tilde{\cal B}_{11}^*\; \omega^2}{\left( \omega^2 - \Omega_0^2\right) + 
\Gamma^2 \:\omega^2}, \nonumber \\
P_Y(\omega, k) &=& \frac{{\cal C}_{Y} + 
\tilde{\cal B}_{22}^*\; \omega^2}{\left( \omega^2 - \Omega_0^2\right) + 
\Gamma^2 \: \omega^2},
\EE
where $\Omega_ {0} = \sqrt{ {\rm det}~{\cal J}^*(k)}$,
$\Gamma = - {\rm tr}~{\cal J}^*(k)$ and where ${\cal C}_{X}(k)$ and 
${\cal C}_{Y}(k)$ are given in Appendix B. Expressions (\ref{eq:pwspec}) 
constitute a full analytical characterization of the spatio-temporal power 
spectra of fluctuations, or equivalently of their spatio-temporal correlation 
properties. This completes the mathematical analysis within the van Kampen 
formalism. The form of the spectra obtained, and their interpretation will 
be discussed in the next section.

\begin{figure}[t]
\begin{center}
\includegraphics[width=70mm]{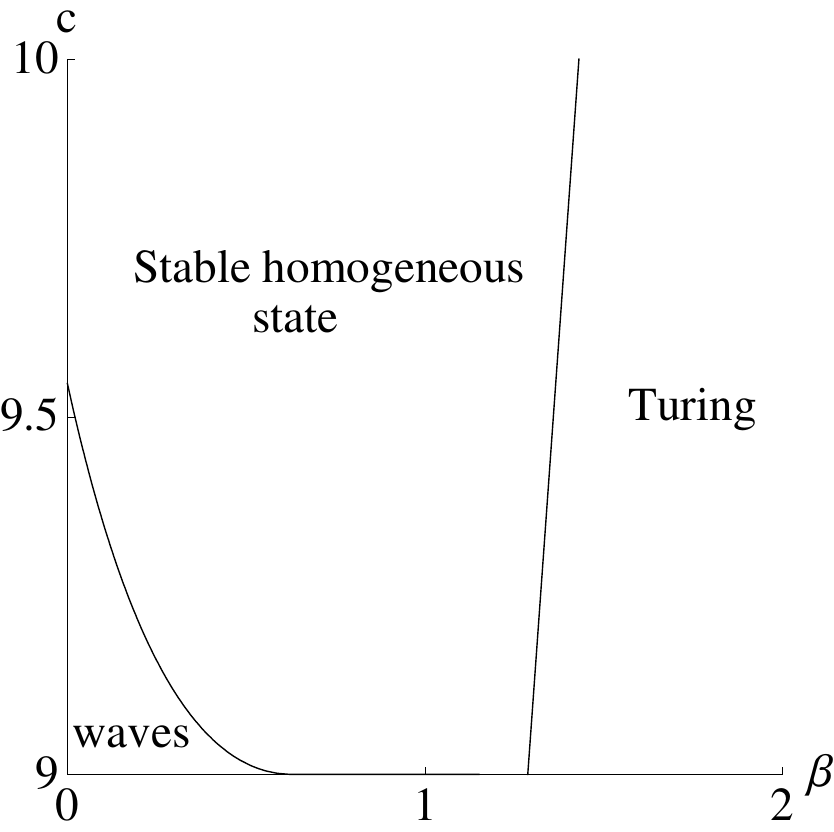}
\caption{Phase diagram in the $(c,\beta)$ plane for the deterministic equations 
\eqref{deter} obtained with a linear analysis of the homogeneous state
\eqref{fps} for $a=d=\alpha=1$, $\sigma=2$ and $b=10$. The model exhibits a 
phase in which the homogeneous fixed point is stable, along with phases with 
Turing patterns and travelling-waves. The two lines in the diagram, $\beta_W(c)$
and $\beta_T(c)$, indicate the onset of these instabilities. Far from these 
lines simultaneous instabilities may occur (e.g.~travelling waves and Turing 
patterns). 
\label{fig:fig1}}
\end{center}
\end{figure}

\section{Results}\label{sec:results}
The main purpose of this paper is to investigate how intrinsic fluctuations 
can bring about stochastic waves. Before we turn to the stochastic dynamics it 
is convenient to characterize the behavior of the nonlocal Brusselator model 
in the deterministic approximation. Having derived analytical expressions for 
the homogeneous state of the deterministic system (see Eq.~(\ref{fps})), as 
well as for the corresponding Jacobian, Eq.~(\ref{Jacobian}), it is 
straightforward to obtain the stability properties of the system as a function 
of the model parameters. This is discussed at the end of Sec.~\ref{sec:model}
and in Appendix A. Given the comparatively large number of parameters, we 
focus on a representative selection of phase diagrams. 

The properties of the deterministic dynamics at varying values of the 
parameters $\beta$ and $c$ (and keeping all other parameters fixed) are
illustrated in Fig.~\ref{fig:fig1}. For a fixed value of $c$ we find a phase 
at intermediate values of $\beta\in[\beta_W,\beta_T]$ in which the homogeneous 
state is stable against fluctuations of any wave number. At a critical value 
$\beta=\beta_T(c)$ a Turing instability sets in, i.e.~an unstable mode occurs 
at a nonzero wave number, with the corresponding eigenvalue being real. At 
values of $\beta$ lower than some second critical value, $\beta_W(c)$, the 
instability occurs again at a non-zero wave number, but now the corresponding 
eigenvalue is complex, indicating a wave instability. We have thus established 
that the nonlocal Brusselator model exhibits Turing instabilities, as well 
as travelling-wave instabilities in the deterministic limit.

\begin{figure}[t]
\begin{center}
\includegraphics[width=70mm]{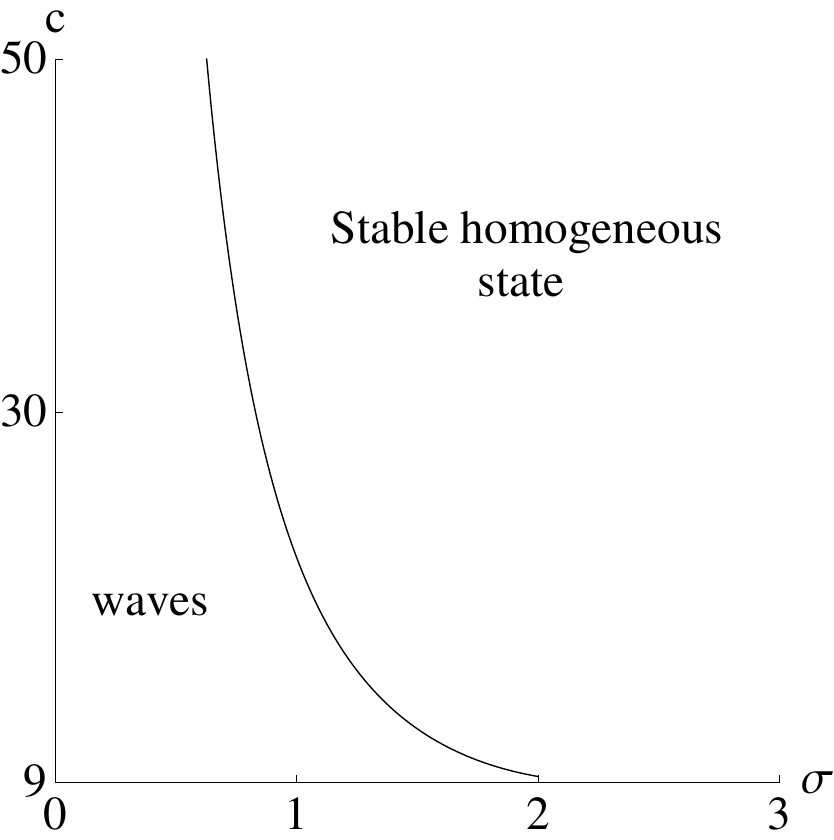} \\
\caption{Phase diagram for the deterministic dynamics, obtained from 
Eqs.~(\ref{deter}) for $a=d=\alpha=1$, $\beta=0.1$ and $b=10$. The solid 
line marks the onset of a wave instability, $c=c_W(\sigma)$. The diagram 
illustrates the role of the nonlocal interaction in generating a wave 
instability. }
\label{fig:fig2}
\end{center}
\end{figure}

In order to illustrate the role of the nonlocal interaction term we show a 
second phase diagram, now in the $(\sigma,c)$-plane, in Fig.~\ref{fig:fig2}. 
Recall that the model parameter $\sigma$ characterizes the range of the 
nonlocal interaction: for small values of $\sigma$ the interaction kernel in 
Eqs.~(\ref{deter}) decays slowly with distance, and the interaction is 
therefore long range. For large $\sigma$ the interaction range is small, in 
the limit $\sigma\to\infty$ one recovers the standard Brusselator model with 
purely local interactions. This is clear from Eq.~(\ref{spacetransprob}); the
only term in the sum which is independent of $\sigma$ is the $j=0$ term, all
the others exponentially decay with $\sigma$, and so vanish as 
$\sigma \to \infty$. Thus the sum tends to $m_{i}/V$ as $\sigma \to \infty$.

As seen in Fig.~\ref{fig:fig2}, the long-range nature of the interactions can 
promote the occurrence of travelling waves in the deterministic system. Varying 
$\sigma$ at fixed value of $c$ (and all other model parameters fixed as 
indicated in the caption of Fig.~\ref{fig:fig2}) one finds that the 
homogeneous state is stable at large values of $\sigma$, i.e. when interactions
are localized. As $\sigma$ is reduced (the interaction range is increased) a 
wave instability sets in.

\begin{figure}[t]
\begin{center}
\includegraphics[width=70mm]{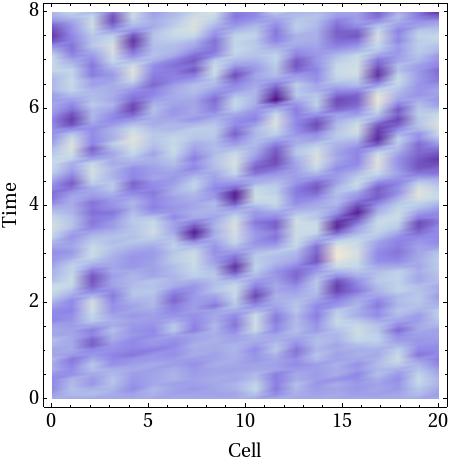} \\
\caption{(Color online) Spatio-temporal dynamics of the concentration of 
$Y$-molecules in the stochastic waves regime. The number of particles, obtained 
from a single run of the stochastic dynamics, is reported on a scale of shades 
(with lighter shades indicating a larger number of $Y_i$-particles per cell, 
and dark shades representing low concentrations). Stochastic waves are seen as 
diagonal structures in the figure, representing both right-moving and 
left-moving waves. The system considered is a one-dimensional ring of $20$ 
cells, each of which has a volume $V = 5000$. Parameter values are 
$a=d=\alpha=1$, $\sigma=2$, $c=b=10$ and $\beta=0.1$.
\label{fig:fig3}}
\end{center}
\end{figure}

\begin{figure}[t]
\begin{center}
\includegraphics[width=70mm]{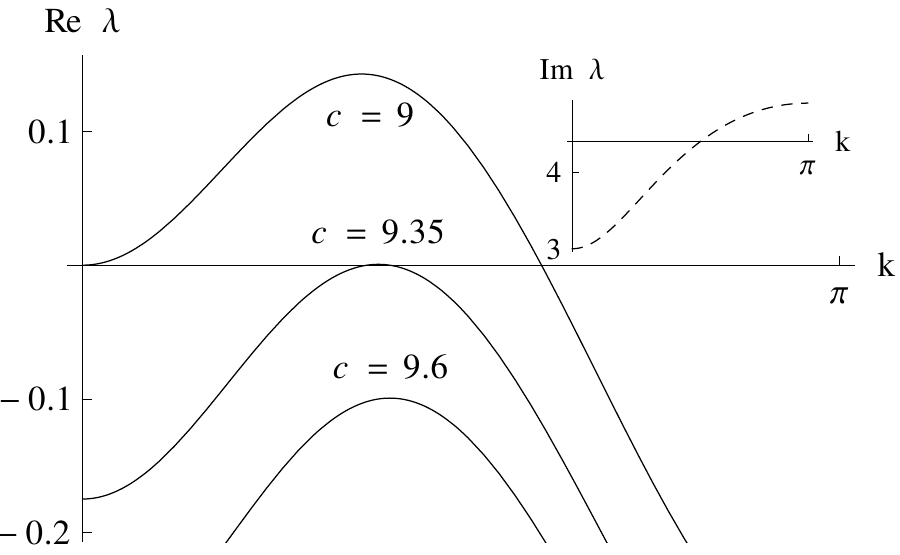}
\caption{Real part (solid line) and imaginary part (inset, dashed line)
of $\lambda(k)$, one of eigenvalues of the Jacobian matrix \eqref{Jacobian} 
(the other is its complex conjugate). Model parameters are 
$a=d=\alpha=1$, $\sigma=2$, $b=10$ and $\beta=0.1$. Using symmetry, we restrict 
the range of $k$ to a half of the Brillouin zone, $[0,\pi]$. The onset of 
the travelling wave instability occurs at $c \approx 9.35$. 
\label{fig:fig4}} 
\end{center}
\end{figure}

We now turn to the stochastic system and show the spatio-temporal behavior of 
the concentration of $Y$-molecules in Fig.~\ref{fig:fig3}. It is important 
to stress that we have chosen values of the model parameters such that the 
deterministic system has a stable homogeneous state. More specifically, for 
the parameters chosen in Fig.~\ref{fig:fig3}, the system is in the stable 
phase of Fig.~\ref{fig:fig2}, but near the line along which a wave instability 
occurs in the deterministic model. As demonstrated by the diagonal structures 
in the space-time representation of Fig.~\ref{fig:fig3}, the stochastic system 
displays travelling stochastic waves for this choice of parameters, even if 
such waves are absent in the deterministic system. In order to illustrate the 
mechanism producing these stochastic waves we plot the dispersion relations of 
modes near the wave instability of the deterministic system in 
Fig.~\ref{fig:fig4} (model parameters other than $c$ are as in the simulations 
shown in Fig.~\ref{fig:fig3}). For $c\gtrsim 9.35$ all modes are stable, and 
the least stable mode (i.e. the mode for which the real part of the 
corresponding eigenvalue is maximal among all modes) has an eigenvalue with 
non-zero imaginary part and negative real part. At $c\approx 9.35$ the wave 
instability line is crossed, one mode is now marginally stable, and the 
corresponding eigenvalue is purely imaginary. Crucially, the instability occurs 
at a non-zero wave number, and with a non-zero imaginary part of the 
corresponding eigenvalue (see inset). For $c\lesssim 9.35$ the deterministic 
system has unstable modes and it exhibits travelling waves. 

\begin{figure}[t]
\begin{center}
\includegraphics[width=70mm]{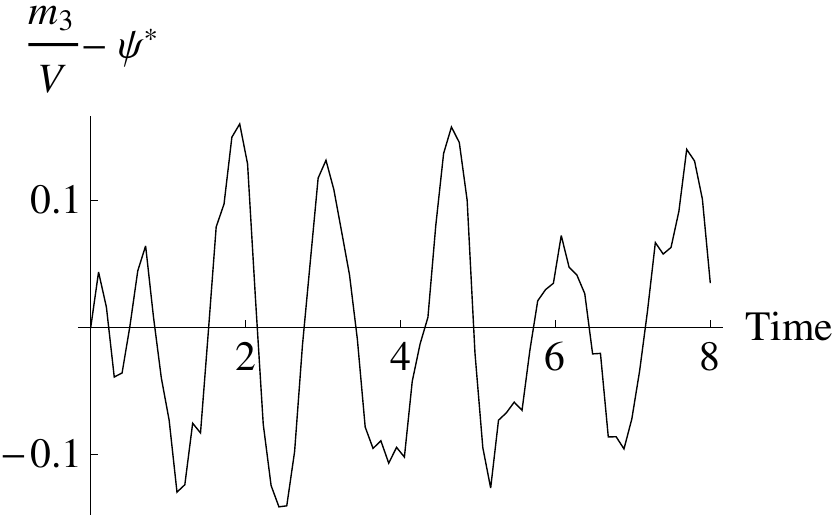}  
\includegraphics[width=70mm]{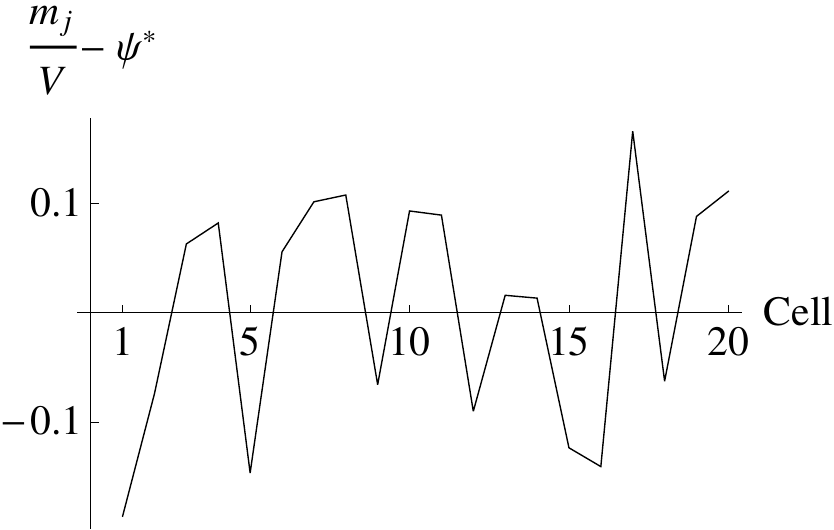} \\
\caption{Upper panel: Temporal evolution of the $Y$-concentration in selected
cell in the stochastic waves regime. Data is obtained from one run of the 
stochastic dynamics, with parameter values as in Fig.~\ref{fig:fig3}. Lower 
panel: Snapshot of the concentration of $Y$-molecules as a function of 
position at a fixed time $t$. 
\label{fig:fig5}}
\end{center}
\end{figure}

The stochastic simulations of Fig.~\ref{fig:fig3} are carried out at $c=10$. 
Here all modes are stable in the deterministic system, the least stable mode  
has a non-zero wavenumber and a complex eigenvalue. In the absence of 
intrinsic stochasticity the system would converge to the deterministic 
homogeneous state. Fluctuations, due to demographic noise at finite cell 
volumes, however constantly cause random motion about the homogeneous state. 
At large, but finite cell volumes, a linear approximation is in order and the 
fluctuations can be decomposed into their Fourier modes. The fluctuations 
corresponding to the least stable mode decay the slowest, on a time scale set 
by the real part of the corresponding eigenvalue. This effect, along with 
fluctuations persistently occurring, conspire due to the mechanism now known as 
coherent stochastic amplification~\cite{McKane2005}. Modes which are stable 
in the deterministic model can effectively be excited by random fluctuations, 
and are observed with sizeable amplitude in the stochastic system. In our 
model this amplitude scales as $V^{-1/2}$ in the cell volume, but the prefactor
multiplying this factor can be significant, resulting in stochastic waves with 
appreciable amplitude even at large cell volumes. 

To illustrate the occurrence of stochastic waves further, we show a time-series 
of the (re-scaled) concentration of Y molecules in a fixed cell in the upper 
panel of Fig.~\ref{fig:fig5}. Coherent stochastic oscillations are clearly 
visible, their amplitude is seen to scale as $V^{-1/2}$ with the cell volume. 
Plotting the time series of the Y-concentration in one cell we have effectively
disregarded the spatial structure of the system, and focused instead on the 
temporal dynamics only. In this sense the upper panel of Fig.~\ref{fig:fig5} 
is similar to observations of amplified stochastic oscillations in the time 
series of non-spatial systems (see for example~\cite{McKane2005}). The lower 
panel of  Fig.~\ref{fig:fig5} shows the concentration of Y molecules as a 
function of position at a fixed time. Spatial modulations, again with an 
amplitude scaling as $V^{-1/2}$, are seen, even though of a lower coherence 
than the temporal oscillations shown in the upper panel. In this lower panel 
we have effectively disregarded the temporal dynamics of the system, and have 
instead focused on its spatial character. The modulations of the concentration 
of Y molecules as a function of position therefore constitutes the analog 
of stochastically-driven Turing patterns reported 
in~\cite{Butler2009, Biancalani2010, Scott2011}. The novelty of our model is 
that it combines the spatial and the temporal aspect; the stochastic waves seen 
in the nonlocal Brusselator model are noise-driven patterns with structure 
both as a function of position and as a function of time.

\begin{figure}[t]
\begin{center}
\begin{tabular}{cc}
\includegraphics[width=70mm]{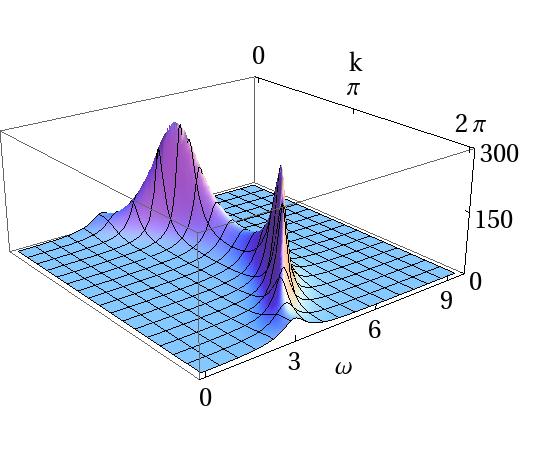} \\
\includegraphics[width=70mm]{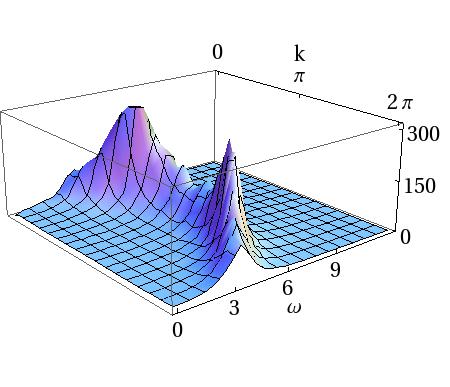} \\
\end{tabular}
\caption{(Color online) Power spectrum of the fluctuations of the $Y$ species 
obtained analytically (upper panel) from Eq.~\eqref{eq:pwspec} and numerically 
(lower panel) by simulating the stochastic process using the Gillespie 
algorithm. The agreement between the two power spectra is clearly very good.
The system is in the stochastic wave regime with  $a=d=\alpha=1$, $\sigma=2$, 
$c=b=10$ and $\beta=0.05$. The spectra show a peaked profile which corresponds 
to spatio-temporal organization despite the deterministic prediction of a 
stable homogeneous state. The numerical spectrum is obtained by averaging 
$200$ realizations of a finite system of $28$ cells each of which has volume
$V=1500$. \label{fig:fig6}}
\end{center}
\end{figure}

The theory developed in Sec.~\ref{sec:stoch_anal} allows us to characterize 
the observed stochastic waves further. Based on the results of 
Eqs.~(\ref{eq:pwspec}) we are able to predict the power spectra of 
spatio-temporal fluctuations about the homogeneous state for any choice of 
model parameters. Stochastic waves are to be expected whenever the highest peak
of these power spectra is found simultaneously at a non-zero wavenumber $k$ 
(indicating non-trivial spatial structure) and at a non-zero angular frequency 
$\omega$ (indicating a complex eigenvalue, and hence oscillatory behavior as 
a function of time). An example of a power spectrum with these properties, 
obtained from the theory of Sec.~\ref{sec:stoch_anal}, is shown in the upper 
panel of Fig.~\ref{fig:fig6}; for comparison we show measurements from direct 
numerical simulations in the lower panel. As seen from this Figure, the 
agreement between theory and simulations is excellent.

\section{Conclusion}
\label{sec:concl}
This paper has been concerned with the investigation of stochastic effects in
a nonlocal extension of the Brusselator model. An analysis of this model on 
the deterministic level reveals that nonlocal interactions can promote the 
occurrence of travelling wave instabilities, similar to what has been seen 
before in other chemical reaction models with long-range 
interactions~\cite{Nicola2006}. As the main result of our work we show that 
the nonlocal Brusselator model can also exhibit travelling waves driven by 
internal fluctuations in parameter regimes in which the deterministic system 
converges to a homogeneous state. Based on a stochastic formulation of the 
model in terms of a chemical master equation, and a subsequent expansion in 
the inverse system size, we derived analytical expressions for the power 
spectra of these spatio-temporal patterns, in excellent agreement with 
direct numerical simulations. 

These findings extend previous results on noise-driven instabilities. 
In~\cite{McKane2005} non-spatial systems were considered, and it was shown 
that intrinsic fluctuations can generate coherent stochastic oscillations 
for parameters for which the deterministic system spirals into a fixed point. 
The work of~\cite{Butler2009,Biancalani2010,Scott2011} instead focused 
on spatial systems, and it was shown that intrinsic noise can generate 
stochastic Turing patterns, i.e.~spatial structures with a constant amplitude 
in time. Our model combines both aspects, and produces stochastic patterns with 
a full spatio-temporal dynamics. 

These stochastic waves are seen in the power spectra of fluctuations, computed 
from the theoretical approach, as isolated peaks at non-zero wave number and 
non-zero angular frequency. While in the past it has been rather difficult 
to observe spatial stochastic patterns directly, and while most of the previous 
work was limited to an indirect identification in Fourier space, we have also 
been able to obtain direct visual confirmation of the stochastic waves 
in the Brusselator model with nonlocal interaction. Criteria which distinguish
stochastic cycles or stochastic patterns from their deterministic analogs 
(limit cycles and Turing-like patterns) have been 
proposed~\cite{Dieckmann2006,Butler2009,Butler2011}. However these are not
applicable to systems with nonlocal interactions of the type we have
considered here, due to the extra $k$-dependence which comes about because 
of the nonlocality. It would be interesting to devise criteria which 
encompass nonlocal models as well.

The purpose of the current work is to illustrate a generic phenomenon which
is expected to occur in a wide class of systems; the Brusselator model was
chosen to illustrate the basic idea because it is simple and widely studied.
The effects of noise on travelling waves has been investigated in the past, but
most of these studies depended on the properties of the specific system under 
consideration. For example, a simple reaction diffusion equation displaying 
travelling waves is the Fisher equation~\cite{Fisher1937}. For this system it 
is known that noise may alter the properties of the waves (see for 
instance~\cite{Moro2001}) and may also be responsible of the emergence of 
noise-driven travelling waves~\cite{Hallatschek2011}. However, these waves 
are different to those we have discussed here, as they do not arise from an 
underlying unstable homogeneous state. A related point is that the Fisher 
equation has only one species, whereas we need at least two interacting 
species, since we require complex eigenvalues to trigger the wave 
instability~\cite{Cross2009}.

We expect that the mechanism of coherent amplification, applies to more 
complicated linear instabilities as well~\cite{Cross2009} and that the concept 
of stochastic waves is relevant in other models with travelling fronts. We also 
expect that the analytical formalism we have developed here for spatial 
systems with nonlocal interaction can successfully be employed to study 
broader classes of individual-based models. This may include models of the 
spread of epidemics, in which infection can occur at a distance; deterministic 
models of such processes have been considered~\cite{Keeling2007,Wang08012010}. 
As stochastic patterns are found in more and more model systems, we expect the 
search for them in real systems to intensify and the understanding of the 
underlying causes of pattern formation in physical, biological and social 
systems to broaden.  

\begin{acknowledgments} 
TB wishes to thank the EPSRC for partial support through a studentship and TG
acknowledges a Research Councils UK Fellowship (RCUK reference EP/E500048/1).
\end{acknowledgments}

\begin{appendix}
\section{Turing and Wave Instabilities}
In this Appendix we derive the conditions which allow the onset of Turing
or wave instabilities to be located. Such criteria exist in the 
literature~\cite{Turing1952,Murray1993,Cross2009}, but most of them rely on 
the standard reaction-diffusion paradigm and will not be applicable in our
case which includes a nonlocal kernel, and so an extra $k$-dependence over and 
above that coming from the diffusion terms. We need therefore a more general 
approach, described below, which closely follows~\cite{Nicola2001}.

We start by defining the region of parameter space in which the homogeneous 
state is stable. Its borders delimitate the instabilities whose type can be 
determined from the analysis below. The stability condition, 
$\mbox{Re}[\lambda_i(k)] <0$ for all $k$, can be conveniently rewritten using 
the trace and determinant if the Jacobian is a $2 \times 2$ matrix as
\be
\mbox{det}\,\mathcal J^* (k) > 0, 
\quad \mbox{tr}\,\mathcal J^*(k) < 0, \quad \forall k.
\ee

The stability region is the set of parameters in which the above inequalities 
hold. Plotting $\mbox{det}\,\mathcal J^*(k_c)$ against
$\mbox{tr}\,\mathcal J^*(k_c)$ we see that we may leave the stability region by
violating one of these inequalities. That is, when
\begin{itemize}
\item There exists a $k_C \ne 0$ such as $\mbox{det}\,\mathcal J^*(k_C) = 0$ 
whereas $\mbox{tr}\,\mathcal J^*(k) < 0 \; \forall k$.
\item There exists a $k_C \ne 0$ such as $\mbox{tr}\,\mathcal J^*(k_C) = 0$ 
whereas $\mbox{det}\,\mathcal J^*(k) > 0 \; \forall k$.
\end{itemize}
It is also possible that determinant and trace become simultaneously zero, but  
this is a degenerate case which we do not consider here. 

Now recall that the eigenvalues of the Jacobian are given by 
\be
\lambda_{1,2} = \frac{1}{2} \left(	\mbox{tr}\,\mathcal J^* \pm 
\sqrt{\left(\mbox{tr}\,\mathcal J^*\right)^{2} 
- 4 \,\mbox{det}\,\mathcal J^* } \right),
\ee
from which the imaginary part of the eigenvalues may be found. From the
discussion at the end of Sec.~\ref{sec:model} we can see that the above 
conditions defining the boundaries of the stability region correspond 
respectively to a Turing instability (the former) and to a wave instability 
(the latter). 

The stability conditions as given are not so convenient to deal with directly, 
because of the presence of inequalities which must be solved for every $k$. To 
overcome this, we suppose that $\mbox{tr}\,\mathcal J^*(k)$ has a global
maximum at $k_M$ and $\mbox{det}\,\mathcal J^*(k)$ a global minimum at $k_m$, a
hypothesis that will be discussed further below. In this case the two 
conditions may be rewritten as:
\begin{itemize}
\item (Turing instability) $\mbox{det}\,\mathcal J^*(k_m) = 0$ and 
$\mbox{tr}\,\mathcal J^*(k_M) < 0$. 
\item (Wave instability) $\mbox{tr}\,\mathcal J^*(k_M) = 0$ and 
$\mbox{det}\,\mathcal J^*(k_m) > 0$.
\end{itemize}
These are the conditions we have used to obtain Fig.~\ref{fig:fig1} and 
Fig.~\ref{fig:fig2}. 

We can now check that the specific forms of $\mbox{tr}\,\mathcal J^*(k)$ and 
$\mbox{det}\,\mathcal J^*(k)$ in our model have the required extrema. Finding
the extremal points of $\mbox{tr}\,\mathcal J^*(k_M)$ can easily be achieved 
analytically; for $\mbox{det}\,\mathcal J^*(k_m)$ it is a little more difficult.
However checking the existence of a global maximum or minimum numerically is
straightforward, and we have verified that for the range of parameters of 
interest to us in this paper such extrema always exist and moreover give the 
boundaries shown in Fig.~\ref{fig:fig1} and Fig.~\ref{fig:fig2}. 

\section{The van Kampen System-Size Expansion}
A description of the general structure and methodology behind the system-size
expansion, as applied to the system under consideration, is given in Section
\ref{sec:stoch_anal}. In this Appendix we give some of the
technical details which would otherwise disrupt the flow of the arguments in
the main text.

The application of the method is facilitated by writing down the master 
equation (\ref{spme}) in terms of the step operators (\ref{sstepop}). An 
example is given in Eq.~(\ref{firstterm}) for the first reaction of the set
of reactions given by Eq.~(\ref{spbrus}). When all eight reactions are 
included the master equation is given by
\BE \label{spme2}
&&{d \over d t}P_{\textbf{n},\textbf{m}}(t) = \nonumber \\
&=&  \sum_{i=-\infty}^{\infty} \Biggl [ (\epsilon_{X,i}^- -1)
\;T(n_i+1, m_i |n_i, m_i) \nonumber\\
&& +(\epsilon_{X,i}^+ -1)\;T(n_i-1 ,m_i, |n_i , m_i) \nonumber\\
&& +(\epsilon_{X,i}^+\:\epsilon_{Y,i}^- -1)\;T(n_i-1, m_i+1 |n_i, m_i) 
\nonumber\\
&& +(\epsilon_{X,i}^-\:\epsilon_{Y,i}^+ -1)\;T(n_i+1, m_i-1| n_i, m_i) 
\nonumber\\
&& +\sum_{j \in \{i-1, i+1\}} \biggl( (\epsilon_{X,i}^+\:\epsilon_{X,j}^- -1)
\;T(n_i-1, n_j+1 |n_i, n_j) \nonumber\\
&& +(\epsilon_{X,j}^+\:\epsilon_{X,i}^- -1)\;T(n_i+1, n_j-1 |n_i, n_j) 
\nonumber\\
&& +(\epsilon_{Y,j}^+\:\epsilon_{Y,i}^- -1)\;T(m_i+1, m_j-1 |m_i, m_j) 
\nonumber\\
&& +(\epsilon_{Y,i}^+\:\epsilon_{Y,j}^- -1)
\;T(m_i-1, m_j+1 |m_i, m_j) \biggr) \Biggr ]\nonumber\\
&& \times P_{\textbf{n},\textbf{m}}(t).
\EE

The fundamental ansatz of the system-size expansion is given by 
Eq.~(\ref{spansatz}) in the main text. It leads to the expression (\ref{vkfs})
for the left-hand side of the master equation. The right-hand side can be
evaluated by observing that the step operators \eqref{sstepop} can be expanded 
in the inverse of the square root of the system size, $V^{-\frac{1}{2}}$, giving 
rise to the following expressions~\cite{Kampen1997}: 
\begin{equation} 
\label{stepdevelop}
\epsilon_{X,i}^\pm \approx 1 \pm \frac{1}{\sqrt V} \partial_{\xi_i} + 
\frac{1}{2V} \partial^2_{\xi_i}, \quad
\epsilon_{Y,i}^\pm \approx 1 \pm \frac{1}{\sqrt V} \partial_{\eta_i} + 
\frac{1}{2V} \partial^2_{\eta_i}.
\end{equation}
This together with the substitution of the ansatz (\ref{spansatz}) into the
transition rates (\ref{spacetransprob}), and the replacement of 
$P_{\mathbf n, \mathbf m}(t)$ by $\Pi (\boldsymbol \xi, \boldsymbol \eta, t )$,
allows us to expand the right-hand side in powers of $V^{-\frac{1}{2}}$. 

Equating the left- and right-hand sides of the master equation gives the 
general form
\begin{equation}
\begin{split}
&\partial_t \Pi 
-\sqrt V \:\nabla_{\boldsymbol \xi}\Pi \cdot \partial_t \boldsymbol \phi 
-\sqrt V\:\nabla_{\boldsymbol \eta}\Pi \cdot \partial_t \boldsymbol\psi \\
& = \left\{ -\frac{1}{\sqrt V} \left[ \boldsymbol f(\boldsymbol \phi,
\boldsymbol \psi) \cdot \nabla_{\boldsymbol \xi} 
+  \boldsymbol  g(\boldsymbol \phi,\boldsymbol \psi) \cdot 
\nabla_{\boldsymbol \eta} \right]  + \frac{\mathbf L}{V} \right\} \Pi,
\end{split}
\label{unscaled_form}
\end{equation}
where $\mathbf L$ is a linear operator containing various derivatives in 
$\eta$ and $\xi$ and $\boldsymbol f$ and $\boldsymbol g$ are functions of
$\boldsymbol \phi$ and $\boldsymbol \psi$. After the introduction of a 
re-scaled time $\tau=t/V$, Eq.~(\ref{gen_form}) of the main text is obtained.
This is now in a form where the various terms on both sides of the equation 
can be balanced.

Equating the leading order terms in Eq.~(\ref{gen_form}) gives equations
whose general structure is displayed in Eq.~(\ref{gen_deter}) and whose 
specific form in Eq.~(\ref{deter}). Equating the next-to-leading order terms
gives an equation with general structure (\ref{fp}) and specific form
\BE
\label{fp2}
\partial_\tau \Pi &=& \sum_{i=-\infty}^{\infty} 
\left( - \sum_{r=1}^2 \partial_{\zeta_{r,i}} 
\left( \mathcal A_{r,i}\Pi \right) \right. \nonumber \\
&& \left. + \frac{1}{2}\sum_{r,s=1}^2 \sum_{j=i-1}^{i+1}  
\partial_{\zeta_{s,i}} \partial_{\zeta_{r,j}} 
\left[ \mathcal B_{rs,ij}\Pi \right] \right),
\EE
where for convenience we have introduced the notation $\zeta_1 \equiv \xi$ and
$\zeta_2 \equiv \eta$. The precise forms of $\mathcal A_{r,i}$ and of
$\mathcal B_{rs,ij}$ will be discussed below, but as mentioned in the main text 
it is easier for our purposes to work with the equivalent Langevin 
equation~\cite{Gardiner1985,Risken1989}
\be
\frac{d \boldsymbol{\zeta}_i}{d\tau} = \mathcal{A}_i(\boldsymbol{\zeta})
+ \boldsymbol{\mu}_i(\tau),
\label{Langevin}
\ee
where $\boldsymbol{\mu_i}$ is a Gaussian noise with zero mean and 
correlator
\be
\langle \boldsymbol{\mu}_i(\tau) \boldsymbol{\mu}_j^{\rm T}(\tau') \rangle =
\mathcal{B}_{ij}\,\delta(\tau-\tau').
\label{noise}
\ee
We are then able to take the Fourier transform of this equation in both space
and time to give Eq.~(\ref{FT_Langevin}).

The Langevin equation~\eqref{FT_Langevin} is defined by two matricies: the 
drift matrix and the diffusion matrix. Since the drift matrix is identical 
to the Jacobian of the system, as presented in Eq.~(\ref{cal_A}), we will
only give the expression for the diffusion matrix.

The diffusion matrix $\mathcal B$ has elements $\mathcal B_{rs,ij}$ where $r$ 
and $s$ index the species and $i$ and $j$ the cell. In the following the 
expressions are given for each $r$ and $s$ and for a given cell $i$. The only
non-zero values of $\mathcal B_{rs,ij}$ occur when $j=i-1, i$ or $i+1$, and
these are given respectively  as the first, second and third entries of a row 
vector:
\BE
&&{\mathcal B}_{11,i} = \Biggl ( -\alpha \Bigl( \phi_i + \phi_{i-1} \Bigr), 
\quad a+(b+d)\:\phi_i \nonumber \\
&& + c \:\Lambda\sum_{j=-\infty}^{\infty} e^{-\sigma |j|}\: \phi_i^2 \psi_{i-j} + 
\alpha \Bigl( \phi_{i-1} + 2 \phi_i + \phi_{i+1}\Bigr), \nonumber\\
&& \quad -\alpha \Bigl(\phi_i + \phi_{i+1} \Bigr) \Biggr ) ,\nonumber \\
&&{\mathcal B}_{12,i} = {\mathcal B}_{21,i} = \nonumber \\
&&= \Biggl (0,\quad -b\phi_i - c \:\Lambda \sum_{j=-\infty}^{\infty} 
e^{-\sigma |j|}\: \phi_i^2 \psi_{i-j},\quad 0 \Biggr), \nonumber  \\
&&{\mathcal B}_{22,i} = \Biggl ( -\beta \Bigl( \psi_i + \psi_{i-1} \Bigr), 
\quad b\:\phi_i \nonumber \\
&& + c \:\Lambda\sum_{j=-\infty}^{\infty} e^{-\sigma |j|}\: \phi_i^2 \psi_{i-j} + 
\beta \Bigl( \psi_{i-1} + 2 \psi_i + \psi_{i+1}\Bigr), \nonumber\\
&& \quad -\;\beta \Bigl(\psi_i + \psi_{i+1} \Bigr) \Biggr ).\nonumber \\
\EE
Evaluating them in the homogeneous state gives
\begin{equation} 
\label{bfp}
\begin{split}
{\mathcal B}^*_{11,i} &=  \left[
\begin{array}{r}
 -\dfrac{2 a \alpha }{d}, \quad
 2 a+\dfrac{2 a b}{d}+\dfrac{4 a \alpha }{d}, \quad
 -\dfrac{2 a \alpha }{d}
\end{array}
\right], \\
{\mathcal B}^*_{12,i} &= {\mathcal B}^*_{21,i} = \left[
\begin{array}{r}
 0, \quad
 -\dfrac{2 a b}{d}, \quad
 0
\end{array}
\right], \\
{\mathcal B}^*_{22,i} &= \left[
\begin{array}{r}
 -\dfrac{2 b d \beta }{a c}, \quad
 \dfrac{2 a b}{d}+\dfrac{4 b d \beta }{a c}, \quad
 -\dfrac{2 b d \beta }{a c}
\end{array}
\right].
\end{split}
\end{equation}

The structure of ${\cal B}_{rs,ij}^*$ can be seen from Eq.~(\ref{bfp}) to be
\be
\label{cal_B_gen}
{\cal B}^{*}_{rs,ij} = b^{(0)}_{rs} \delta_{i-j,0} + b^{(1)}_{rs} \delta_{|i-j|,1},
\ee
where the two matrices $b^{(0)}$ and $b^{(1)}$ can be read off from 
Eq.~(\ref{bfp}). It is then straightforward~\cite{lugo2008} to calculate the 
spatial Fourier transform, $\tilde{\cal B}_{rs} \equiv \tilde{\cal B}_{rs}(k)$, 
of the matrices \eqref{bfp} with respect to the variable $i-j$:
\be
\tilde{\cal B}_{rs}^*(k) = (b^{(0)}_{rs} + 2 b^{(1)}_{rs}) + 
b^{(1)}_{rs}\tilde\Delta,
\ee
where $\tilde{\Delta}$ is given by Eq.~(\ref{e_and_Delta}). The explicit 
forms are
\BE
\tilde{\cal B}_{11}^*(k) &=& \frac{2a}{d}\left( b + d \right) 
- \frac{2a\alpha}{d}\tilde{\Delta}, \nonumber \\
\tilde{\cal B}_{12}^*(k) &=& \tilde{\cal B}_{21}^*(k) = - \frac{2ab}{d},
\nonumber \\
\tilde{\cal B}_{22}^*(k) &=& \frac{2ab}{d} 
- \frac{2bd\beta}{ac}\tilde{\Delta}. 
\label{FT_cal_B}
\EE

The power spectra of the stochastic oscillations, defined by 
Eq.~(\ref{PS_defn}), have the form (\ref{eq:pwspec}). The functions
${\cal C}_{X}$ and ${\cal C}_{Y}$ are defined by
\BE 
\label{pwstuff}
{\cal C}_{X}(k) &=& \tilde{\cal B}_{11}^*(k) \:{\cal J}_{22}^*(k)^2 
- 2 \:\tilde{\cal B}_{12}^*(k)\: {\cal J}_{12}^*(k)\: {\cal J}_{22}^*(k) 
\nonumber \\ 
&& + \tilde{\cal B}_{22}^*(k)\: {\cal J}_{12}^*(k)^2, \nonumber\\ \nonumber \\
{\cal C}_{Y}(k) &=& \tilde{\cal B}_{22}^*(k)\: {\cal J}_{11}^{*}(k)^2 
- 2\: \tilde{\cal B}_{12}^*(k)\: {\cal J}_{21}^*(k)\: {\cal J}_{11}^*(k) 
\nonumber \\
&&+ \tilde{\cal B}_{11}^*(k)\: {\cal J}_{21}^{*}(k)^2.
\EE

\end{appendix}

\end{document}